\documentclass[preprint,10pt,nocopyrightspace]{sigplanconf}
\usepackage{comment}
\usepackage{epsfig}
\usepackage{multirow}
\usepackage[obeyspaces,hyphens]{url}
\usepackage{color}
\usepackage{times}
\usepackage{array}
\usepackage{ragged2e}
\usepackage{ifthen}
\usepackage{pifont}
\usepackage{amsmath, amssymb}
\usepackage{tabularx}
\usepackage{microtype}
\usepackage{booktabs}
\usepackage{endnotes}
\usepackage{paralist}
\usepackage{wrapfig}
\usepackage{xspace}
\usepackage[small,compact]{titlesec}
\usepackage{dcolumn}
\usepackage{verbatim}
\usepackage{etoolbox}
\usepackage{natbib}

\usepackage{balance}
\usepackage{graphicx}
\usepackage{listings}
\usepackage{color}

\makeatletter
\AtBeginEnvironment{thebibliography}{%
   \interlinepenalty=10000}
\makeatother

\definecolor{dkgreen}{rgb}{0,0.6,0}
\definecolor{gray}{rgb}{0.5,0.5,0.5}
\definecolor{mauve}{rgb}{0.58,0,0.82}

\newboolean{longconferencenames}
\setboolean{longconferencenames}{false}

\ifthenelse{\boolean{longconferencenames}}
{

}{

}

\makeatletter
\patchcmd{\ttlh@hang}{\parindent\z@}{\parindent\z@\leavevmode}{}{}
\patchcmd{\ttlh@hang}{\noindent}{}{}{}
\makeatother

\usepackage[
bookmarks,bookmarksopen,
pdfstartview={FitH},
colorlinks,linkcolor={black},citecolor={black},
urlcolor={black}
]
{hyperref}
\usepackage{etoolbox}

\newboolean{publicversion}
\setboolean{publicversion}{true}
\ifthenelse{\boolean{publicversion}}{
  \newcommand{\grumbler}[2]{}
}{
  \newcommand{\grumbler}[2]{\textcolor{red}{\xspace\bf #1: #2\xspace}}
}

\newcommand{\ttt}[1]{{\tt #1}\xspace}

\newcommand{\paragraphbe}[1]{\vspace{0.75ex}\noindent{\bf \em #1}\hspace*{.3em}}

\newcommand{\sys}{Chiron\xspace}
\newcommand{\Sys}{\sys}
\newcommand{\worker}{training enclave\xspace}
\newcommand{\workers}{training enclaves\xspace}
\newcommand{\Worker}{Training enclave\xspace}
\newcommand{\Workers}{Training enclaves\xspace}
\newcommand{\server}{parameter server\xspace}
\newcommand{\Server}{Parameter server\xspace}
\newcommand{\oracle}{query enclave\xspace}

\newcommand{\spcode}{service provider code\xspace}
\newcommand{\Spcode}{Service provider code\xspace}

\title{\Sys: Privacy-preserving Machine Learning as a Service\vspace{30pt}}

\newcommand{\ut}{The University of Texas at Austin}
\authorinfo{Tyler Hunt}{\ut}{}
\authorinfo{Congzheng Song}{Cornell University}{}
\authorinfo{Reza Shokri}{National University of Singapore}{}
\authorinfo{Vitaly Shmatikov}{Cornell Tech}{}
\authorinfo{Emmett Witchel}{\ut}{}
\begin{document}
\maketitle
\subsection*{Abstract}

Major cloud operators offer machine learning (ML) as a service, enabling
customers who have the data but not ML expertise or infrastructure to
train predictive models on this data.  Existing ML-as-a-service platforms
require users to reveal all training data to the service operator.

We design, implement, and evaluate \sys, a system for privacy-preserving
machine learning as a service.  First, \sys conceals the training data
from the service operator.  Second, in keeping with how many existing
ML-as-a-service platforms work, \sys reveals neither the training
algorithm nor the model structure to the user, providing only black-box
access to the trained model.

\sys is implemented using SGX enclaves, but SGX alone does not achieve
the dual goals of data privacy and model confidentiality.  \sys runs the
standard ML training toolchain (including the popular Theano framework
and C compiler) in an enclave, but the untrusted model-creation code from
the service operator is further confined in a Ryoan sandbox to prevent
it from leaking the training data outside the enclave.  To support
distributed training, \sys executes multiple concurrent enclaves that
exchange model parameters via a parameter server.

We evaluate \sys on popular deep learning models, focusing on benchmark
image classification tasks such as CIFAR and ImageNet, and show that its
training performance and accuracy of the resulting models are practical
for common uses of ML-as-a-service.

\section{Introduction}

The impressive accuracy achieved by modern machine learning (ML)
in business~\cite{businessML}, medicine~\cite{nvidiaMed}, and
communication~\cite{skype,googletranslate} motivates many data holders
to apply ML to their own datasets.  Existing ML frameworks, however,
are not easy to deploy by non-expert users due to a large number of
configuration parameters and general lack of understanding of why and
how modern ML works.  Furthermore, ML expertise is scarce and often
unrelated to data holders' primary competency (e.g., customer relationship
management or biomedical research).

This mismatch has created a growing business in \emph{machine learning
as a service}, with offerings from Google~\cite{googleMLaaS},
Amazon~\cite{aml}, and Microsoft~\cite{msMLaaS}, and several
startups~\cite{bigml,mljar,havenondemand,nexosis}.  ML-as-a-service
promises to bring high-quality ML techniques to non-expert users,
but privacy and confidentiality are obstacles to its adoption.
Machine learning is useful insofar as it enables accurate quantitative
predictions.  When using conventional ML services, users are revealing
precisely the data that (they hope) can give them a competitive advantage.
If the training data is sensitive\textemdash commercial transactions,
confidential documents, medical images, etc.\textemdash owners will not
want to expose it to ML service providers and thus will not be able to
take advantage of their services.

On the service provider side, many commercial ML services, including
Google's Prediction API and Amazon ML, do not reveal their training
algorithms or resulting models to the customers.  In part, this helps
support their business: they give customers API access to the trained
models and charge per API use.  They may also be concerned about the
intellectual property contained in their proprietary models and the
configuration parameters they use for particular ML tasks.  We assume that
ML services are not actually interested in stealing their customers' data
and would be willing to forego their access to this data in exchange for
the increased adoption of their services by privacy-sensitive customers.

\paragraphbe{Our contributions.}
We present \sys,\footnote{In Greek mythology, Chiron is a centaur
entrusted with training demigods and heroes.} a system that enables data
holders to train ML models on an outsourced service without revealing
their training data.

The service provider is free to choose the type of the model to train,
how to configure and train it, and what transformations, if any,
to apply to the inputs into the model.  These choices can adaptively
depend on the user's data and ML task.  The user obtains API access to
the trained model but no other information about it.  This matches how
ML-as-a-service operates today.

To enforce data confidentiality while allowing the provider to
select, configure, and train a model any way they want, \sys employs a
Ryoan~\cite{ryoan} sandbox, which in turn is based on a hardware-protected
enclave such as Intel's SGX~\cite{sgx}.  An enclave alone is insufficient
because it only protects \emph{trusted} code executing on an untrusted
platform.  Code can only be trusted if it is public and thus can be
checked by users.  In \sys, however, the ML service provider's code
is \emph{untrusted}, thus users must be assured that this code is not
stealing their data even though they cannot inspect it.

\Sys leverages the confinement provided by Ryoan to enable the service
provider's code to access users' data, then define and train a model,
while preventing it from exfiltrating the data.  Users can verify that
the the enclave is executing a Ryoan sandbox extended with standard ML
toolchain code, but without seeing the specifics of the model being
trained.  Ryoan inherits SGX limitations, such as vulnerability to
cache-timing and memory-access channels (discussed in \S\ref{s:sgxbroke}).

For performance, \sys places a generic, model-indep\-end\-ent ML
toolchain\textemdash in particular, the standard Theano framework and
C compiler\textemdash inside the hardware-protected enclave but outside
the sandbox.  The untrusted code inside sandbox can only interact with
this framework via an interface controlled by the sandbox and cannot
use it to leak information. The integrity of the toolchain and sandbox
is attested to the user.  Therefore, the user only needs to trust
the sandbox and ML toolchain, both of which are public and standard.
There are no checks specific to the ML model.

Modern ML models, especially deep learning models, benefit from concurrent
training of multiple models on different batches of the training data.
\sys provides support for concurrent training via a parameter server and
data-oblivious channels between enclaves that enable them to exchange
model parameters during training.

To evaluate \sys, we focus on the popular deep learning approach that
demonstrated exceptional accuracy for many classification tasks.
We measure the convergence times, scalability, and performance
of \sys when training ML models on standard benchmarks such as
CIFAR~\cite{krizhevsky2009learning} and ImageNet~\cite{imagenet}.
We also explore the effects of training multiple models and of varying
the parameter exchange rate for several levels of parallelism.


\section{Background}\label{s:back}

\label{s:ml-background}

\subsection{Machine learning}

A \emph{machine learning model} (ML model) is a function with a set of
parameters that maps an input to a target output.  For example, the inputs
to a facial recognition model are images, the target is the identity
of the person in the image.  The model parameters are usually floating
point numbers.  For example, a linear regression model is a function
$f(x) = W^\top x$ where $W$ is the parameters (aka the weight vector)
and $x$ is the input.  In artificial neural networks, the parameters
are the weights on the connections between the nodes of the network.

We focus on \emph{supervised learning}.  The \emph{training data} in this
case is a set of input points, each labeled with its correct target.
The goal of training is to find a set of parameters that maximizes the
model's accuracy on the training data.  This is usually done by optimizing
the \emph{objective (or loss) function}, which penalizes the model when
it outputs a wrong target on a data point.  A \emph{regularization} term
can be added to the objective function to penalize model complexity.
This helps prevent overfitting, when the model exhibits high accuracy
on the training data but poor accuracy on the test data.

After training has finished, the model is evaluated using \emph{test
data} which was not used during training.  A standard metric is \emph{test
accuracy}, which is the percentage of test data points that are classified
correctly.

For many ML tasks and models, the objective function cannot be optimized
in one step.  Instead, the learning algorithm initializes the model
parameters, then iteratively feeds batches of training data to the model,
calculates the loss, and updates the parameters to reduce the loss.
Training typically finishes when the loss value stops decreasing or when
parameter updates become smaller than a certain threshold, i.e., when
the model \emph{converges}, or else after a fixed number of iterations,
or when the model achieves acceptable accuracy on a validation dataset
(distinct from the training dataset).

Most modern ML algorithms have a set of tunable \emph{hyper-parameters},
distinct from the model parameters.  Hyper-parameters control the
configuration of the algorithm such as the number of training iterations,
the ratio of regularization term in the loss function, the size of each
training batch, etc.

\paragraphbe{Deep learning.}  
Deep learning has become very popular for many ML tasks,
especially related to computer vision and image recognition
(e.g.,~\cite{krizhevsky2009learning,krizhevsky2012}).  Deep learning
models are composed of layers of nonlinear mappings from input to
intermediate hidden states and finally to output.  Each connection between
layers has a floating point weight matrix as parameters.  These weights
are updated during training.  The topology of the connections between
layers is task-dependent and important to the ultimate accuracy of
the model.

Training a deep neural network on a large dataset can take a long time.
A common way to scale deep learning is through data and model parallelism,
with multiple models training concurrently and exchanging parameters
via a parameter server~\cite{dean2012, chilimbi2014, shokri15}.
The training dataset is partitioned into subsets and a separate model
is trained on each subset.  The parameter server stores the global set
of all model parameters.  During training, each local model pulls the
parameters from this server, calculates the updates based on its current
batch of training data, then pushes these updates back to the server,
which updates the global parameters.

\paragraphbe{Model design vs.\ model training.} In the architecture of
\sys, we distinguish \emph{model design} from \emph{model training}
(see \S\ref{s:worker}).  Model design code specifies the type and
topology of the model, the loss function, the optimization algorithm,
the values of the hyper-parameters, and the transformations, if any, to
apply to the model inputs (e.g., scaling, rotating, or resampling images).

Model training is the generic process of repeatedly applying the model
to a batch of data, calculating the loss, and updating the parameters
according to the specifications provided by the model design code.

\subsection{Symbolic computation}

Most ML algorithms are essentially series of mathematical operations,
e.g., matrix multiplication that maps input to output or the loss
calculation given the model's output and the corresponding target values.
These operations can be defined symbolically and thus training algorithms
can be constructed independently of the data.

Modern ML libraries including Theano~\cite{theano},
TensorFlow~\cite{tensorflow}, Chainer~\cite{chainer}, and
MXNET~\cite{mxnet} support symbolic computation.  A model is defined
as a computational graph with input and output nodes.  Inner nodes are
mathematical operations, such as element-wise addition and dot product,
that connect the computation in order from input to output.  Data is fed
into the input node and flows through the series of operations defined
by the graph.

\label{sec:theano_example}

Figure~\ref{fig:lr_build} shows a simple example that defines a logistic
regression model in Theano for a binary classification task.  The goal
is to classify input $x$ as belonging to class $0$ or class $1$.
The model is defined mathematically as $y=\sigma(W^\top x)$ where $W$
are the parameters we are trying to learn and $\sigma(z) = \frac{1}{1
+ e^{-z}}$ is the sigmoid function with range $[0,1]$. The output $y$
models the probability of $x$ being classified as $1$.

\begin{figure}[t]
\begin{lstlisting}
def logistic_regression(dim, lr):
	# initialize parameters as all zeros
	W = theano.shared(value=np.zeros(dim))
	# input and target node
	x = T.matrix('x')
	t = T.vector('t')
	# sigmoid function on dot product
	# between input and W
	y = T.nnet.sigmoid(T.dot(x, W))
	# loss function
	loss = 	T.nnet.binary_crossentropy(y, t).mean()
	# gradient of parameters w.r.t loss
	grad_W = T.grad(loss, W)
	# stochastic gradient descent update
	sgd_step = {W: W - lr * grad_W}
	# building computation graph
	train_func = theano.function([x, t], loss, updates=sgd_step)
	return train_func
\end{lstlisting} 
\caption{Defining logistic regression in Theano.}
\label{fig:lr_build}
\end{figure}

The loss function for logistic regression is cross-entropy, which
is defined as $l(y, t) = -(t\log(y) + (1 - t)\log(1 - y))$, where
$y$ is the output defined above and $t$ is the true target label
(0 or 1). Overall, for training data $D=\{(x_i, t_i)\}_{i=1}^n$,
our objective is to minimize $L = \frac{1}{n}\sum_{i=1}^n l(y_i, t_i)$.
The minimization problem can be solved using stochastic gradient descent,
which calculates the gradient of the model parameters with respect to loss
$L$ and updates the parameters by subtracting the gradients multiplied
by the \emph{learning rate} hyper-parameter.

Figure~\ref{fig:lr_build} shows how this model is defined in Theano.
It first initializes parameters $W$, then defines the input and
target nodes of the computational graph, then defines the loss and the
parameter update functions.  Finally, it calls \ttt{theano.function}
to build the computational graph given the input data nodes, output
loss node, and parameter updates.  Internally, Theano will optimize the
graph and generate and compile C code for each operation in the graph.
\ttt{train\_func} is a callable object which takes data as input,
computes the loss for that data, and updates the parameters accordingly.
During training, a batch of training data is fed into \ttt{train\_func}
in each iteration until the model converges.

\begin{figure}[t]
\begin{lstlisting}
def design_model(dim1, dim2, lr):
	# initialize parameters as all zeros
	W1 = theano.shared(value=np.zeros(dim1, dim2))
	W2 = theano.shared(value=np.zeros(dim2))
	# input and target node
	x = T.matrix('x')
	t = T.vector('t')
	# Layer1: relu function on dot product
	# between input and W1 
	h = T.nnet.relu(T.dot(x, W1))
	# Layer2: sigmoid function on dot product
	# between hidden layer and W2
	y = T.nnet.sigmoid(T.dot(h, W2))
	# loss function
	loss = 	T.nnet.binary_crossentropy(y, t).mean()
	# gradient of parameters w.r.t loss
	grad_W1, grad_W2 = T.grad(loss, [W1, W2])
	# stochastic gradient descent update
	sgd_step = {W1: W1 - lr * grad_W1,
				W2: W2 - lr * grad_W2}
	# building computation graph
	train_func = theano.function([x, t], loss, updates=sgd_step)
	return train_func
\end{lstlisting} 
\caption{Defining a simple neural network in Theano.}
\label{fig:nn_build}
\end{figure}

Figure~\ref{fig:nn_build} shows another Theano example, defining a
two-layer neural network for a binary classification task.  The model
is $y=\sigma(W_2^\top \text{relu}(W_1^\top x))$ where $W_1$ and $W_2$
are the parameters we want to learn and $\text{relu}(z) = \max(z, 0)$
is the activation function.  The loss function and parameter updates
are the same as in the logistic regression example.

\subsection{Machine learning as a service}
\label{sec:mlservice}

ML-as-a-service platforms provide convenient APIs for users to upload
their data and train an ML model.  The trained model can be returned
directly to the user or made available for querying through a special
API.  Many major cloud providers now offer this service, including
Google's Prediction API~\cite{googleMLaaS} (soon to be replaced by Cloud
Machine Learning Engine), Amazon ML~\cite{aml}, and Microsoft's Azure
ML~\cite{msMLaaS}.

ML-as-a-service APIs are usually provided as black boxes.  In many
services, the user does not know the type of the model selected by the
provider (which could depend on the user's data and task) or the details
of the training.  Google's Prediction API hides all details; users have no
information about how the model is designed and trained.  Amazon ML lets
users choose a few hyper-parameters such as model size, regularization,
and the number of training iterations.  The choice of the model depends
on these hyper-parameters but is invisible to the user.  Model training
involves stochastic gradient descent but the implementation details are
hidden.  Microsoft's Azure ML provides a wide range of built-in models.
Users can choose a model but have no information about the implementation
details of the learning algorithm.

Our design of \sys preserves this separation between model design, which
is proprietary to the service operator and not available to the user,
and model training, which is a generic procedure of repeatedly applying
the training function to batches of training data.

\subsection{Hardware-protected enclaves}
\label{s:back-sgx}

Software Guard Extensions (SGX~\cite{sgx}), available on Intel processors
starting with Skylake, provide \emph{enclaves} that protect code and data
from all other software on the platform, including privileged software
such as the operating system and hypervisor.  Code in an enclave can
safely operate on secret data without fear of unintentional disclosure
to the platform.  The privacy and integrity of the enclave is enforced
by hardware~\cite{sgx}.  Enclaves are sometimes called trusted execution
environments (TEEs), but we will use ``enclave'' as a generic term.

SGX supports remote attestation of the code and data that make up the
initial state of the enclave.  This enables a remote user to verify that
the initial code and data matching a given cryptographic hash are loaded
into a genuine enclave.  Remote attestation is always the first step in
bootstrapping a secure channel to an enclave.

\paragraphbe{Attacks on SGX.}
The design of SGX leaves untrusted privileged software in control
of system resources, enabling it to mount subtle attacks (see
\S\ref{s:sgxbroke}) which are beyond the scope of this paper.

\paragraphbe{Rollback.}
A platform can rollback persistent state.  While SGX allows enclaves
to hash and encrypt data for storage using a hardware-generated secret
key, it provides no guarantees about freshness.  This drawback forces
enclave designers to rely on hardware counters~\cite{ariadne}, which
have limitations and performance issues, or software-based strategies
that leverage other machines~\cite{rote}.

\paragraphbe{Enclave indistinguishably.}
While SGX enables enclaves to attest their integrity to outside parties,
nothing prevents the platform from instantiating multiple copies of
enclaves.  Without a mechanism to uniquely identify different instances
of the same enclave, a malicious platform could confuse remote users about
the state of a particular enclave by maliciously redirecting communication
between a set of enclaves that remote users perceive as a single enclave.

\subsection{Ryoan}

Ryoan~\cite{ryoan} enables service providers to keep proprietary code
secret while simultaneously ensuring users that the confined code cannot
leak their data.  Instead of asking users to trust the provider's code,
Ryoan asks them to trust the sandbox that confines this code.  Users can
audit Ryoan to gain confidence in its correctness.

Ryoan is based on Native Client~\cite{nacl,nacl-64}, which uses compiler
techniques to confine code.  Binaries are checked at load time to ensure
they are properly restricted. Confined code relies on the sandbox for
all interactions with the outside world.  Ryoan runs inside an SGX
enclave, which removes the need to trust the privileged software of the
computational platform.


\section{Threat Model}
\label{s:threat_model}

The primary goal of \sys is to protect users' data from malicious
providers of ML-as-a-service.  We assume that training a model and
deploying it afterwards both take place on the provider's computational
platform.  Training data should remain confidential while the model is
being trained, even if the model architecture, loss and optimization
functions, and the hyper-parameters of the training algorithm are defined
by the service provider.  Queries to the model and its outputs should
remain confidential when a trained model is being used.

We assume that the entire platform is untrusted, including the privileged
code such as the operating system and hypervisor.  The attacker could
be the machine's owner and operator, a curious or even malicious
administrator, or an invader who has taken control of the OS and/or
hypervisor.  The attacker may own a virtual machine (VM) physically
co-located with the VM being attacked or she could even be a malicious
OS developer and add functionality that directly records user input.
Therefore, \sys aims to prevent the untrusted code used during training
from exfiltrating secrets about the training data to the underlying
platform.

By default, \sys does not reveal any information about the model
architecture or training to the user.  This matches the current practice
of many commercial ML-as-a-service operators (see \S~\ref{sec:mlservice}).

\paragraphbe{Trusted computing base.}  
Both the user and the service operator must trust Ryoan's sandboxing code,
which we assume will be distributed by a disinterested third party.
Ryoan is a generic sandbox that is not tailored to machine learning.
Our prototype makes several modifications to Ryoan to accommodate \sys,
such as the logic for distributed enclaves to coordinate model setup
(see \S~\ref{s:design-overview} for a summary of these modifications).

Ryoan requires both parties to trust the hardware and its implementation
of SGX and SGX is more complicated than most hardware
features~\cite{boumann17hotos}.  Remote attestation for enclaves
requires trusting a remote attestation service~\cite{sgx_rem_attest}. In
\S\ref{s:limits}, we discuss the limitations of \sys with respect to the
covert and side channels affecting SGX.

We assume a standard, generic ML toolchain for
defining and training models.  In our prototype, we use the popular
Theano framework for this purpose.  Because the code of ML toolchains
such as Theano is public and well-understood, we assume that it has been
scrutinized and that it does not deliberately steal information about
the training data and exfiltrate it from the enclave where it executes.
The enclave attests to the identity of Ryoan, and Ryoan attests to the
identity of the ML toolchain (further details about the chain of trust
are available in~\cite{ryoan}).

\paragraphbe{Denial of service.}  
Denial of service is outside the scope of our threat model.  The service
provider's model design code can simply refuse to run or the underlying
untrusted operating system can refuse to schedule our trusted code.
The provider can skip the training or train an inaccurate model.
We assume that the user will test any generated model with data held
back from the training dataset to see if the model is sufficiently
accurate for their purposes.  This is identical to the current usage
of ML-as-a-service.

\paragraphbe{Adversarial access to the model.}
A trained model accessed by an adversary, even via black-box queries, may
leak its training data either accidentally~\cite{shokri2017membership},
or because the model was maliciously designed~\cite{song2017ccs} to
reveal its training data in response to certain queries.  This appears
to be a generic feature of ML models, such as deep neural networks,
that have very high memorization capacity.  This threat is still poorly
understood and no generic mitigations are currently available (see
discussion in \S\ref{s:related}).

\sys loads trained models into special query enclaves protected with a key
that the provider does not know (see \S\ref{s:query}).  A model can be
queried only by the user whose data was used to train it.  Adversaries,
including the provider itself, can neither observe the user's queries,
nor query the model themselves.  This scenario corresponds to many common
uses of ML-as-a-service.

\section{Design}\label{s:design}
\begin{figure}
\centering
\includegraphics[width=\columnwidth]{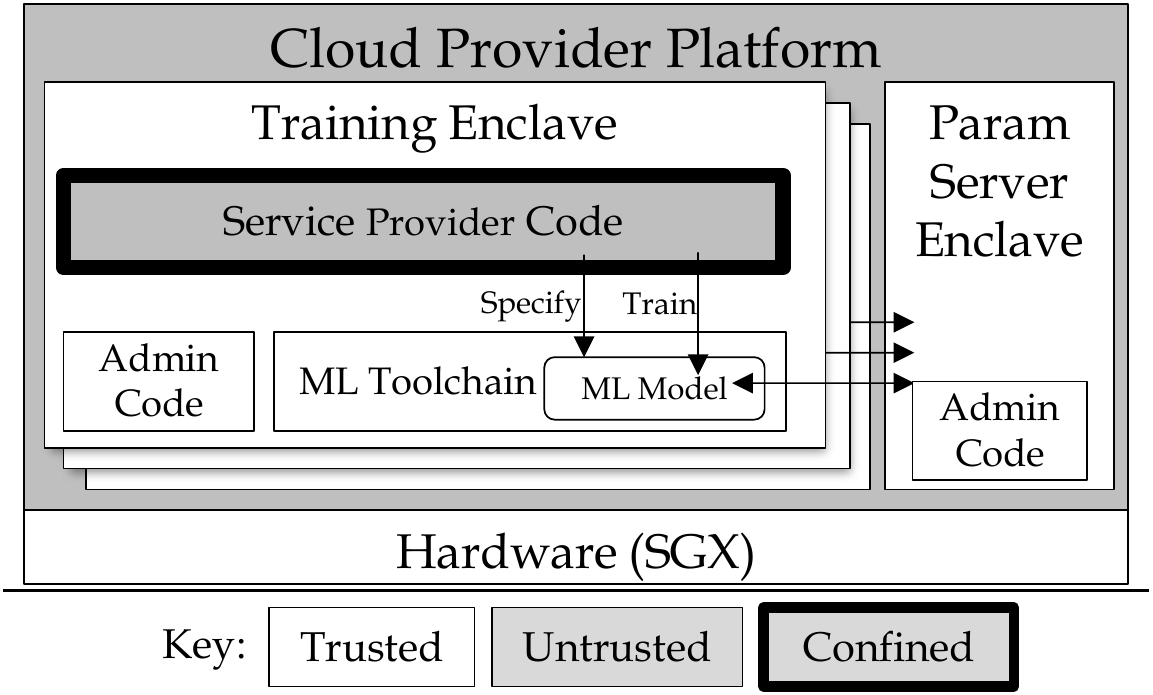}
\caption{\Sys architecture.}
\label{fig:arch}
\end{figure}

\label{s:design-overview}
The main component of \sys is the \emph{\worker} (seen in
Figure~\ref{fig:arch}), which consists of a Ryoan sandbox (which is part
of ``Admin code'' in the figure) extended with a standard ML toolchain
(Theano, in our implementation).  The cloud provider platform encompasses
all system software including the operating system and hypervisor.
The code in the \worker is public, so its integrity can be remotely
attested using standard remote attestation for enclaves~\cite{sgx}.
The service provider loads its own untrusted code into the Ryoan sandbox
and makes one or more \workers available to the user.  The user connects
to \workers and submits data.  \Spcode then performs two tasks.

First, \spcode sets up the model architecture, loss function, optimization
function, and training hyper-parameters.  These choices may depend on
the user's data, but \sys confinement prevents \spcode from leaking
information about that data outside the sandbox.

Some existing ML services allow users to specify certain hyper-parameters.
For example, Amazon ML lets users set the number of training iterations.
In principle, \sys could support this by adding trusted enforcement
code to the ML toolchain that checks the service provider's compliance
with the user-specified hyper-parameters.  In our current prototype,
we assume that the service provider controls all hyper-parameters (this
matches how Google's Prediction API operates today).

\Spcode must use \sys's ML toolchain to define the model; all other forms
of output are disallowed by \sys's confinement.  The ML toolchain compiles
the model description into executable, model-specific training code
inside the enclave.

Second, \spcode drives the training of the model by feeding user data to
the ML toolchain and invoking the training code that updates the model
parameters.  If necessary, \spcode may apply \emph{data transformations}
(e.g., scaling or rotating images) before passing the data to the
training code.  The ML toolchain executes this code as usual, but the
service provider cannot observe the state of the model due to SGX and
Ryoan sandbox protections.

When the training is distributed over multiple \workers, each working
on its own shard of the user's training data, \sys coordinates them
using a dedicated \server enclave.  Each \worker periodically pulls
model parameters from the \server, updates them, and sends them back
to the \server.  This helps all \workers collectively converge to the
same model.  \Workers exchange parameter updates with the \server via
secure channels that only send fixed-sized messages at a fixed rate.
Updates thus do not leak information about the training data or the
current state of the model.

Figure~\ref{fig:arch} shows the architecture of \sys.  \Spcode is confined
in several \workers, where it may interact through a controlled interface
with a trusted ML toolchain.  All enclaves contain trusted admin code,
which includes the code for establishing secure channels, managing access
to other outside resources, and loading and sandboxing \spcode.

After training has finished, \sys outputs a model encrypted with a key
known only to the user.  The service provider may send the model to the
user or keep it, in which case the model is instantiated in a dedicated
query enclave and the user must use \sys to submit encrypted queries
and receive the model's outputs.

\paragraphbe{Ryoan modifications.}
First, \sys replaces Ryoan's communication model.  Ryoan is designed for
request-oriented services, allowing untrusted code to send one message
for every request.  \Sys is optimized for the iterative nature of ML
training: it allows many messages for a single input, but sends them at
a fixed rate.

Second, Ryoan supports labeled data to help multiple service providers who
do not trust each other to avoid disclosing their secrets to the user.
This functionality is not useful for \sys, which assumes that all
untrusted code is written by a single service provider.

Finally, Ryoan mitigates the cost of constructing enclaves by providing
a mechanism to reuse initialized enclaves without mixing data from
different users.  This is important to applications that run on the
order of seconds per request.  \sys, however, is designed to support
ML model training, which runs on the order of hours or even days, and
initialization time is a tiny fraction of the total execution time.
Therefore, \sys does not reuse enclaves because the time and space
overheads for reuse outweigh the benefits.

\subsection{Initialization} \label{s:init}

To initialize \sys, the service provider starts up the \server and a set
of \workers.  The \server and \workers attest each other and construct
secure communication channels (see \S\ref{s:appenc} for details).
Having established a secure channel with each \worker, the \server
generates a nonce using randomness from the processor and sends it to
the \workers over the secure channels.  This nonce uniquely identifies
the \server instance.

After receiving the nonce, \workers load untrusted \spcode and wait to be
contacted by the user.  The user verifies the integrity of each \worker
via attestation and establishes a pairwise secure channel with each one
(see \S\ref{s:appenc} for details).

Through secure channel establishment, each \worker learns the user's
public key.  This public key is forwarded to the \server and the
\server's nonce is forwarded to the user.  To ensure that all \workers
have been contacted by the same user, the server checks whether it
received the same public key from every \worker; otherwise, it does not
allow training to begin.  The user ensures that all \workers are using
the same \server by comparing the nonces received from the \workers;
otherwise, he does not send the data.  These checks are necessary since
SGX does not prevent the platform from instantiating multiple, unexpected
copies of \sys enclaves and directing network traffic to them.

The user then shards training data into $n$ pieces for $n$ \workers
(randomly or according to any other criterion) and sends each shard
to a distinct \worker along with the learning task (e.g., if the task
is classification, the specification includes the output classes of
the model).  Each \worker exposes the data shard and the learning task
to \spcode so that it can define the model and begin training.

\begin{figure}
\centering
\includegraphics[width=\columnwidth]{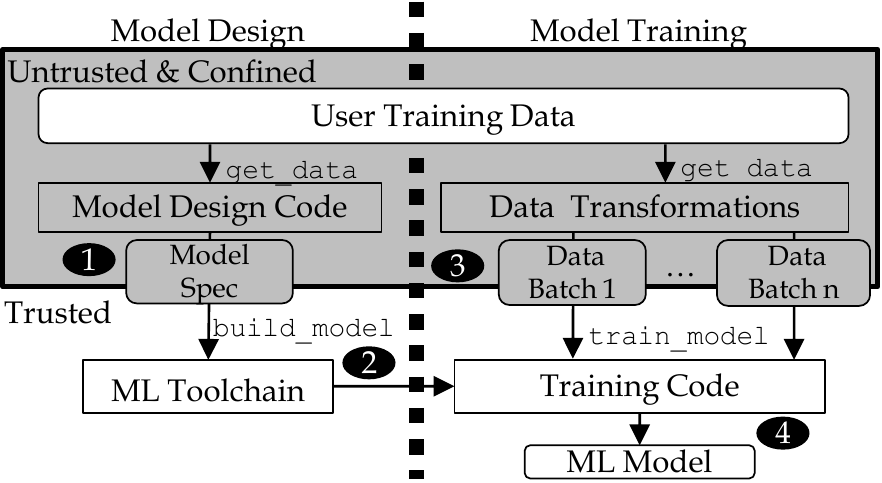}
\caption{\Worker architecture. 1) Untrusted \spcode examines data, then
generates a model spec and passes it to the ML toolchain. 2) ML toolchain
uses the spec to generate model-training code. 3) \Spcode transforms
data and breaks it into batches for training. 4) Model-training code is
invoked for each batch, updating the model.}
\label{fig:worker}
\end{figure}

\subsection{\Worker}
\label{s:worker}

Each \worker is a Ryoan~\cite{ryoan} sandbox augmented with an
ML toolchain, as depicted in Figure~\ref{fig:worker}.  The sandbox
confines \spcode and supports the following interfaces: \ttt{get\_\-data}
allows confined code to read user data, \ttt{build\_\-model} allows it
to construct a model, \ttt{train\_\-model} allows it to train the model
for a single iteration.  The following interfaces are not depicted in
the figure to avoid clutter: \ttt{examine\_\-model} allows untrusted code
to examine the model as it is being trained, \ttt{test\_\-model} allows
it to test the model on validation data, and \ttt{finish} notifies the
trusted code that training is complete (unless the user gets to set the
number of iterations).

Each \worker has a life cycle consisting of four stages: initialize, model
design, model training, and dead.  Transitions between the stages are
determined by the confined code's use of the interfaces described above.

\paragraphbe{Initialize stage.}
Each new \worker starts in this stage.  It initializes trusted code before
loading the untrusted \spcode.  In this stage, \spcode can interact with
the network and persistent storage, which it may use to initialize itself
(e.g., load the latest model configurations).  After this process has
finished, the untrusted code calls \ttt{get\_\-data} to transition to
the model design stage.  If data is not yet available, the \worker will
wait until it is available before returning control to the \spcode.

\paragraphbe{Model design stage.}
In the model design stage (shown in the left half of
Figure~\ref{fig:worker}), the untrusted \spcode has access to the
user's data and learning task.  \Spcode is now confined and its
access to the network and persistent storage is cut off by the
admin code in the \worker.  Confined code may call \ttt{get\_\-data}
at any time to read some or all of the data.  The specification of the
learning task is always prepended to the data and \spcode can access it
through \ttt{get\_\-data}.  This allows \spcode to adaptively choose the
architecture and hyper-parameters of the model depending on the nature
of the data or the user-specified task.  For example, for an image
classification task, it may adaptively choose a support vector machine
(SVM) or an artificial neural network.

\begin{figure}[t]
\begin{lstlisting}
def untrusted_design_model(dim1, dim2, lr):
	# same design code as in Figure 2
	...
	# serialize the python objects
	model_str = serialize(x, t, loss, sgd_step)
	# make a call to trusted build_model and 
	# let Chiron take over
	build_model(model_str)
	


\end{lstlisting} 
\caption{Building a model in \sys.}
\label{fig:modelbuild}
\end{figure}

\Spcode defines the model, including its computational graph, loss
function, and the hyper-parameters.  These are standard Theano operations,
illustrated in Figures~\ref{fig:lr_build} and~\ref{fig:nn_build} in
Section~\ref{sec:theano_example}.

In \sys, model design is separated from model training.  After defining
the model, \spcode must call \ttt{build\_\-model} with a description
of the model, as shown in Figure~\ref{fig:modelbuild}.  This invokes
generic, data- and model-independent Theano code, which is trusted and
runs outside the sandbox.  It generates the model-training function.
The model is now ready for training and the \worker transitions to the
model training stage.

\paragraphbe{Code generation.}
Modern ML libraries such as Theano optimize the training process
by generating, compiling, and executing model-specific native code.
While it is technically possible to build and train an ML model without
code generation, this dramatically decreases training performance.
Therefore, in \sys we opted to keep code generation intact.

Our sandboxing mechanism~\cite{ryoan} relies on examining all code
before it is loaded.  Therefore, we cannot let \spcode directly generate
executable code lest it use this code to escape the sandbox.  Instead,
we use a trusted Theano toolchain to generate the code (the same way
as it does in conventional usage), based on the specification from the
sandboxed \spcode.

An alternative solution would be to include the ML toolchain in \spcode
and provide a sandbox interface that (1) checks that the generated code
does not violate the sandbox and (2) safely loads it.  This would result
in a smaller trusted computing base because \sys would no longer trust
the code generation component of the ML toolchain.  On the negative side,
this approach would incur a non-trivial performance penalty.  For example,
the authors of Native Client report execution time overheads for sandboxed
code as high as 43\% on SPECC2000 benchmarks~\cite{nacl-64}.  We leave
a thorough exploration of this approach to future work.

\paragraphbe{Coordination between \workers.}
Because each \worker receives a different shard of the data, \sys allows
confined code to broadcast one message of a pre-defined size to all
other \workers.  The purpose of this message is for all \workers running
confined model design code to agree on the same model.  For example, the
service provider can designate one \worker as a leader who defines the
model based on its sample of the user's data (or, alternatively, based
on the reports from other \workers) and broadcasts its decision to all
\workers. The timing of this message creates creates a timing channel
(see \S\ref{s:limits} for details), thus this added flexibility comes
at the cost of leaking a few more bits.

If this additional leakage is unacceptable, users could randomly select
a data sample and send it to all \workers along with the data shards.
This enables the service provider to coordinate model choice without
exchanging messages if this choice is deterministic.

\paragraphbe{Model training stage.}
The model training stage is shown in the right half of
Figure~\ref{fig:worker}.  Before returning control to \spcode, the \worker
contacts the \server, initializing copies of the newly created model
parameters.  If they already exist, the \worker instead updates local
parameters with the current values obtained from \server.  Throughout
the training stage, a separate thread exchanges periodic updates with
the \server.  After exchanging data with the \server, \sys returns
control to \spcode.  \Spcode updates the local parameters by calling
\ttt{train\_model} on a batch of training data.  The \ttt{train\_model}
function may be called as many times as necessary.

Each invocation of \ttt{train\_\-model} applies \ttt{train\_\-func}.  This is
a callable object, generated from the model definition, which takes a
batch of data as input, computes the loss for that data, and updates
the parameters accordingly. The \ttt{train\_func} object is generated
by the trusted ML toolchain and is therefore trusted not to leak.

\Spcode signals to trusted code that training has finished by calling
\ttt{finish}.  \Spcode may call \ttt{examine\_\-model} and \ttt{test\_\-model}
at any time during model training to help decide that the model is ready.
The \ttt{examine\_\-model} interface allows \spcode to examine properties
of the model (e.g., the values of the loss function or parameter
gradients in the most recent round of training).  The \ttt{test\_\-model}
interface allows untrusted code to evaluate the model on validation data.
This approach gives service providers the most flexibility.  \Sys could
easily adopt other policies, e.g., letting the user specify the total
number of \ttt{train\_\-model} invocations.

\paragraphbe{Dead stage.}
After transitioning to the dead stage, \spcode is no longer executed.
\sys encrypts the final model with a symmetric key provided by the user.
It then returns a hash of the model to the user, before storing the
encrypted model in the platform.  See \S\ref{s:query} for an explanation
of how the user can access the model.

\subsection{Parameter exchange}
\label{s:param-exch}

\paragraphbe{\Server.}
The \server is a trusted, in-memory key-value store which runs in a
dedicated enclave.  \Workers initialize a particular set of keys that
represent model parameters.  During training, the \server exchanges
parameters with each \worker.  This design is standard in distributed
machine learning, e.g., it is similar to the parameter store in Project
Adam~\cite{chilimbi2014}.

Updates from each \worker contain the differences between its current
parameters and those it received in the last communication with the
\server.  The \server applies these differences to its own copies of the
parameters and replies with the updated values.  The \worker replaces
its parameters with the updated values from the \server.

\paragraphbe{Fixed-rate exchange.}
Exchanges between each \worker and the \server are network operations
and therefore visible to the platform.  The content of the messages is
always encrypted but \sys does not hide the fact that communication is
taking place.

\Sys prevents \spcode from turning these exchanges into a covert channel
by ensuring that their timing is data-independent.  Instead of tying the
frequency of exchanges to the frequency of updates (e.g., exchange after
every call to \ttt{train\_model}), \sys adopts a fixed-rate update policy.
The parameter-exchange thread never synchronizes with the data-processing
thread.  Instead, it sleeps for a configurable amount of time, performs
an exchange, then sleeps again.

\Sys relies on the platform to schedule the parameter-exchange thread
according to this policy.  Relying on the platform as a source of time
is safe in this case because the platform cannot see the training data,
thus any scheduling policy it chooses must be data-oblivious.

\subsection{Secure communication channels}
\label{s:appenc}

Communication channels between the enclaves and between the user and the
enclaves are secured using AES-GCM~\cite{gcm}, and an authenticated key
agreement protocol, which together provide end-to-end application-level
encryption, message integrity, and endpoint authentication.  Keys
are derived separately for each connection and never persisted.
Per-connection keys keep the platform operator from transparently
replacing enclaves because a new enclave will need to establish a new
channel, alerting the victim to the disruption.

To establish communication with a new enclave instance, \sys uses the
standard, SGX-specific Diffie-Hellman key exchange protocol as defined
by Intel~\cite{sgx-guide}.  This protocol uses hardware-signed SGX
attestations to identify enclaves and prevent man-in-middle attacks.
Each attestation uniquely identifies the initial code and data of the
enclave being attested, thus this protocol allows only valid enclaves that
are running the correct software to exchange keys.  Every message in the
key exchange is embedded in an attestation and thus signed by hardware.
Enclaves use true hardware-generated randomness from the processor
to initialize the secret parameters (i.e., private Diffie-Hellman
exponents) of the exchange.  The user is identified by a public key
rather than hardware attestation; other than this, enclave-to-enclave
and user-to-enclave key exchange protocols are identical.

In principle, \sys could include an implementation of transport layer
security (TLS).  Adopting TLS for inter-enclave communication, however, is
an overkill because SGX attestations replace TLS authentication and these
attestations are already required by \sys to verify enclave integrity.
Inter-enclave communication does not need to be backward-compatible
with other code.  When contacting the user, \sys uses TLS because \sys
does not directly control the code on the user's machine.  Furthermore,
TLS certificates may be useful to verify the origin of untrusted \spcode.

\subsection{Using a trained model}
\label{s:query}

\Sys enforces model ownership: only the owner of the data used to train
the model may query this model.  The model is encrypted with a key provided
by the user, and \sys provides a ``\oracle'' that the service provider can
run to safely decrypt the model and respond to user queries to the model.
Each \oracle contains enough of the ML toolchain to load the trained
model, apply it to user's query, and return the output to the user.

\Sys stores trained models in platform controlled persistent storage,
which is vulnerable to rollback attacks (see \S\ref{s:back-sgx}).
To ensure that the user always uses the freshest model, the user attaches
the expected hash of the model to any query, and \sys ensures that the
hash matches the loaded model before responding to the query.

We assume that after the model has been created, the user measures its
test accuracy on a subset of the data withheld from the training dataset
(see \S\ref{s:ml-background}) and proceeds to use the model only if test
accuracy is acceptable.

As mentioned in \S\ref{s:design-overview}, using the model may require
applying some transformations to the inputs.  These transformations may
be proprietary to the service provider and thus cannot be disclosed to
the user.  To support this functionality, \sys includes a Ryoan sandbox
that confines untrusted data transformation code written by the service
provider and loaded into the \oracle.  Untrusted provider code within
the \oracle can examine and query the model.  The result of the query is
returned to the untrusted code, which then passes it to the user over
the secure channel.  Since the untrusted code is confined, it cannot
leak this result to the service provider.  \Sys pads or truncates the
message to a fixed size to ensure that message sizes are data-oblivious.


\section{Implementation Issues}
\label{s:impl}

\paragraphbe{SGX compatibility.}
All \sys components are designed to run in SGX enclaves.  We use an SGX
compatibility layer implemented in libc.  Our modified libc marshals
system call arguments and passes them to the training enclave; it also
unmarshals results.  Finally, it protects against all currently known
Iago attacks~\cite{iago}.

\paragraphbe{C compiler.}
\sys must replace Theano's C compiler, which is gcc.  To avoid a side
channel from the compilation process, the compiler running in the \worker
cannot write files, because that would leak model-building activities
to the untrusted platform.  We could not get gcc to stop writing files,
thus \sys uses libclang and llvm's execution engine to compile the
model training code (following the ``clang-interpreter'' example from
the llvm repository~\cite{llvm-jit}).  We measure the performance cost
of this decision in Section~\ref{s:eval-cifar}.

\paragraphbe{\Server.}
The \sys \server is based on Redis version 3.2.8~\cite{redis}, modified
to support \sys's application-level encryption and to apply received
updates to the in-memory parameters.  \Workers initialize a key for each
model parameter.  In each parameter exchange, the \server adds any value
received from the \workers to the value of the associated key and returns
the updated value.

The \server uses time from the platform to calculate the epoch (e.g.,
\texttt{nanosleep}).  The platform can deny service by providing
inconsistent time but cannot leak the data via a timing channel (see
\S\ref{s:param-exch}).

\paragraphbe{Managing data within enclave.}
\Sys requires that users shard their data before uploading it to the
set of \workers.  Our \sys prototype assumes that each shard will fit in
enclave memory because this simplifies the design.  In our experiments,
the shards are always small enough to fit.  That said, enclave memory
is a limited resource, with current SGX hardware restricted to 128MB
of enclave memory.  Furthermore, physical memory for enclaves must be
statically partitioned at boot time, and the memory dedicated to enclave
use is not usable by non-enclave code.  This static partitioning will
likely lead to conservative partition sizes.

When data does not fit into memory, \sys streams data in.  Trusted code in
each \worker reads a chunk of data of configurable size at a configurable
fixed rate (similar to our fixed-rate parameter exchange), overwriting
the old chunk.  It is the service provider's job to ensure that the
fixed rates are adequate given the time it takes to train on each chunk.
We leave exploration of efficient policies for managing insufficient
enclave memory for future work.


\section{Limitations}
\label{s:limits}

\paragraphbe{Covert and side channels.}
\label{s:sgxbroke}
\sys is based on Ryoan, which blocks untrusted code from exfiltrating
information about the training data to the platform using software
interfaces such as syscall sequences and arguments.  \sys inherits Ryoan's
limitations regarding covert and side channels, which in turn result from
the limitations of the underlying hardware\textemdash in particular,
the internal monitoring that an untrusted platform can perform using
the processor monitoring unit (PMU).  Intel explicitly places certain
timing channels outside the scope of SGX~\cite{sgx-side-channels}.
Untrusted code can leak bits by modulating cache accesses, page accesses,
execution time, etc.  These limitations are discussed at length by Hunt
et al.~\cite{ryoan}.

The current specification for SGX allows privileged software to
manipulate the page tables of an enclave to observe its code and data
trace at page-level granularity.  This can lead to devastating attacks
that use application-specific information to reconstruct fine-grained
secrets from these coarse addresses, e.g., words in a document and
images~\cite{xu15oakland}.  While these channels pose a serious risk
for SGX as it exists today, we hope that hardware vendors will mitigate
them in the future.  Even assuming the current SGX architecture, there
is active ongoing research on how to use other processor features, e.g.,
transactional memory support, to detect and prevent privileged software
from attacking enclaves~\cite{chen17accs,shih17ndss}.

In a machine learning context, one software technique for mitigating
page-based side channel attacks is to transform the learning algorithm
so that it is data-oblivious, i.e., its access pattern is independent
of the training data~\cite{ohrimenko2016}.  \sys, however, is designed
to support ML-as-a-service where service providers do not reveal their
models to users.  In this setting, the provider's code that sets up the
model is both untrusted and hidden from the user.  There is no way for
the user to verify that it is indeed data-oblivious.

\sys also has \sys-specific timing channels.  For example, an adversarial
service provider can encode some secret about the training data in
the number of training iterations (if that hyper-parameter is under
the provider's control), time until the model coordination message is
broadcast or the training time. Similarly, refusing
to terminate or outputting a deliberately inaccurate model is another
channel. Our current prototype does not mitigate these channels which can
leak dozens of bits given the current design of \sys.  Standard defenses
include quantizing or padding execution time.

\paragraphbe{Lack of GPU support.}
State-of-the-art deep learning models rely on GPUs to achieve high
performance during training~\cite{krizhevsky2012}.  \sys cannot use GPUs
because \sys fundamentally relies on hardware-supported trusted execution
environments, such as those enabled by SGX.  These environments do not
exist on today's GPUs.  If users' data is processed on a GPU, we are
not aware of any techniques that can protect it from the GPU operator.

Some frameworks for distributed deep learning~\cite{dean2012,chilimbi2014}
do not use GPUs because GPUs' memory limitations inherently restrict
the size of the models that can be trained.  More recent work showed
how to take advantage of model parallelism~\cite{ShazeerMMDLHD17} and
specialized parameter servers~\cite{cui2016geeps} to support distributed
learning with GPUs.

\section{Evaluation}
Our evaluation of \sys is conducted on two machines (A and B). Machine A
has a 8-core Intel Xeon 3.20GHz processor with 16 GB RAM and machine B
has a 4-core Intel Core i7-6700 3.40GHz processor with 32 GB RAM.
We use Theano version 0.8.2. The Ryoan sandbox used is based on
NaCl commit 2d5bba1. All enclaves use Ryoan's modified eglibc version
2.19.

\textbf{CIFAR} is an object classification dataset with 50,000
training images (ten categories, 5,000 images per category) and 10,000
test images~\cite{krizhevsky2009learning}.  Each image is 32x32 pixels,
each pixel has three 8-bit values corresponding to RGB intensities.

To benchmark the CIFAR dataset, we use a 9-layer VGG-style
network~\cite{vgg} with batch normalization~\cite{batch_normalization}.
The trained model has 2.3 million parameters.  We use L2 regularization
with the ratio of 0.0005 and stochastic gradient descent with the learning
rate of 0.1, batch size of 128 and 60 training epochs (each epoch sweeps
through the whole dataset).

\textbf{ImageNet} is a large-scale visual object recognition
dataset~\cite{imagenet}.  The original dataset has over 1.2 millions
images in 1,000 categories.  We picked a subset with 250,000 images
in 200 categories (ImageNetLite), which is a reasonably large dataset
for current ML-as-a-service applications.  We held out 25,000 images
for test evaluation.  It takes about 38 hours to train a model on this
dataset, which is large enough to represent real workloads while making
experimentation tractable.

We use AlexNet for ImageNetLite training, with the same preprocessing and
configuration as~\cite{krizhevsky2012}.  We scale each image to 128x128
pixels with three color channels (RGB) and train the network on a randomly
cropped 112x112 image.  During evaluation, we deterministically crop the
image to 112x112, following the same procedure as the AlexNet authors.
The model has 6 million parameters.  We also use L2 regularization with
the ratio of 0.0005, stochastic gradient descent with the learning rate
of 0.05, batch size of 64, and 60 training epochs.

\subsection{Performance modeling}
\label{s:sgx-perf}
The \sys prototype requires features only supported by SGX V2.  At the
time of writing there is no publicly available processor which supports
SGX V2. SGX V2 introduces additional instructions which can change
memory mappings of running enclaves. In accordance with
Ryoan~\cite{ryoan} (and other related work~\cite{haven14osdi,vc3}), we
evaluate performance using an SGX performance model that mixes
measurements from an SGX V1 processor with reasonable guesses for
latencies of V2 instructions.

Our SGX performance model includes penalties for enclave exit events,
the major source of overhead introduced by SGX. Events that cause enclave
exits are page faults, system calls, and interrupts, which all flush the TLB.
Exiting and reentering the enclave also comes with
additional overhead from executing SGX instructions.  System calls
require an explicit enclave exit then reenter (EEXIT, EENTER); page
faults and interrupts trigger an involuntary (asynchronous in the
literature~\cite{sgx}) exit and require a resume operation (ERESUME).

We measure SGX instruction overheads on real SGX hardware (a Dell Inspiron
7359 laptop with Intel Core i5-6200U 2.3 GHz processor). Measurements
were collected using Intel’s SGX Linux Driver~\cite{sgx-driver}
and SDK~\cite{sgx-sdk}.  Explicit exits incur a $3.9$ microsecond
penalty, involuntary exits incur a $3.14$ microsecond penalty. We use a
modified Linux kernel that counts each exit event and flushes the TLB for
designated enclave threads.  We multiply the exit counts by the overheads
measured on hardware and add the result to the enclave execution time.

As explained in \S\ref{s:impl}, \sys marshals and copies system call
arguments and return values across the enclave boundary, which also adds
execution time overhead.

\subsection{Parameter server throughput}

Machine learning is computationally intensive and benefits from being
split across multiple enclaves, which exchange parameter updates via
a \server.  The \server receives updates (usually 4-byte floating point
numbers) and performs a single floating point addition for each received
delta (applying it to the corresponding parameter).  The \sys prototype
\server is based on Redis version 3.2.8~\cite{redis}.  All operations are
done in memory; all persistence mechanisms provided by Redis are disabled.
When using dummy clients that flood the \server with updates, it saturates
at 22.6 million floats per second (about 86MB/s), confirming that the
network hardware and stack is not a bottleneck for our experiments.
While this performance is sufficient for our experiments, \server
implementation could be improved.  Redis is single-threaded, which is
limiting when it must also perform floating point operations.

\subsection{CIFAR experiments}
\label{s:eval-cifar}

\begin{table} \small
  \centering
\begin{tabular}{r | l | r |  r} 
\toprule
   $n$ & Ex Rate & Test Acc(\%) & Time (hr)\\
\hline
& Baseline & 89.56 & 9.70 \\
& Ideal & 89.38 & 10.07\\
2 & 1/3 avg & 89.17 & 9.92 \\
& 1/2 avg & 88.50 & 9.79 \\
& Avg & 88.98 &  9.69  \\
\hline
& Baseline & 88.89 & 6.09 \\
& Ideal & 88.97 & 6.86 \\
4 & 1/3 avg & 88.38 & 6.59 \\
& 1/2 avg & 87.66 & 6.70 \\
& Avg & 88.07 & 6.74 \\
\hline
& Baseline & 88.52 & 3.14 \\
& Ideal & 88.28 & 3.79 \\
8(1) & 1/3 avg & 86.66 & 3.91 \\
& 1/2 avg & 86.24 & 3.81 \\
& Avg & 81.09 & 3.73 \\
\hline
& Baseline & 88.41 & 3.18 \\
& Ideal & 88.05 &  3.68 \\
8(2) & 1/3 avg & 86.90 &  3.57 \\
& 1/2 avg & 86.46 & 3.67  \\
& Avg & 84.63 & 3.66 \\
\end{tabular}

   \caption{
   Performance and accuracy with different parameter exchange policies
   and number of training enclaves ($n$).  Baseline is unmodified Theano,
   exchanging parameters after each training epoch.  Ideal is Chiron
   without fixed-rating the exchange.  Avg is Chiron with the fixed
   rate set to the average duration of one training iteration. Half and
   one-third Avg exchange at twice and three times the rate
   of Avg.  For 8 enclaves, the 8(1) configuration runs all 8 enclaves on
   machine A (8 cores), 8(2) runs five enclaves and the parameter server
   on machine A and the remaining three enclaves on machine B (4 cores).
   \label{tab:cifar} }
\end{table}
We compare \sys to the conventional baseline training time and accuracy
while varying the number of training enclaves from two to four to eight
($n$).  Table~\ref{tab:cifar} shows the training time, including
latencies from the SGX performance model (\S\ref{s:sgx-perf}), for
different parameter exchange policies (Ex Rate).
The table also shows the test accuracy of the resulting models, as
measured by the percentage of successfully classified test data.

The table shows a drop in test accuracy for the resulting model as
the number of enclaves increases.  This is a common observation in
the asynchronous distributed learning setting, known as the gradient
staleness problem~\cite{staleness}.  The problem arises because by the
time a worker has finished one round of calculating updates and has
sent the results to the parameter server, the parameters may have been
updated a number of times.  Very stale updates disrupt the convergence
of the model. Fixing this issue is an active research problem in the
machine learning community.  By fixed-rating the communication between
the parameter server and training enclaves, test accuracy can decrease
by up to 6\%.  Higher rates of fixed-rate communication result in more
accurate models.

In the ideal case, local parameters are downloaded from the parameter
server right before a training iteration starts, and the updates are sent
right after the iteration finishes. When communication is fixed rate,
there is some time between downloading the parameters and starting an
iteration (similarly there is time between sending updates and finishing
the iteration).  During this lag, the parameters could be updated which
would result in updates that are more stale than the ideal case.

The performance of \sys using the ideal parameter exchange policy lags the
baseline by 4--20\%.  Most of this overhead is during the computationally
intensive training phase that executes within the enclave but outside
the Ryoan sandbox. Overheads come from enclave exits that cause TLB flushes,
slowing the memory-intensive training process.  90\% or more of the
enclave exits are from page faults, not asynchronous sources, so they
are difficult to eliminate.  The performance of training does not change
much with different parameter exchange policies.

\subsection{ImageNetLite experiments}

\begin{table} \small
   \centering
   \begin{tabular}{l|r|r|r|r}
      \toprule
      & Top1(\%) & Top5(\%) & Train(hr) & Query(sec)\\
      \hline
      Baseline & 55.12 & 78.51 & 39.83 & 3825.30\\
      \hline
      Chiron & 52.41 & 76.42 & 38.85 & 3843.53
   \end{tabular}
   \caption{Model accuracy, training time, and query time for ImageNetLite.
     Top 1 is the accuracy for the most likely prediction.
     Top 5 is the accuracy of the five most likely predictions.
     Query shows time for querying 100,000 images in batches of 1,000.
     }  \label{t:macro}\label{f:query}
\end{table}
We demonstrate \sys's ability to scale to more substantial ML tasks by
training ImageNetLite using 16 \workers. We use the one-third Avg parameter
exchange policy in this experiment, which does not leak training time and
as shown by the CIFAR experiments, provides best
model accuracy.  Table~\ref{t:macro} shows that \sys slows down
ImageNetLite training by 16\%, while preserving the accuracy of the
trained model.  For comparison, a random guess with five chances would
have a top-5 accuracy of 0.025.

\Sys must be invoked in order to perform queries on the model. Table~\ref{f:query}
shows the performance impact \sys imposes on model queries, which is less than 1\%.

\paragraphbe{Cost of outsourced training.} 
While \sys supports training on machines owned by the service provider,
this is not a requirement.  At the time of this writing, Amazon EC2
charges \$0.0665 per core per hour for computation and \$0.02 per GB
for network communication (between data centers)~\cite{ec2-numbers}.
Training ImageNetLite on the baseline would cost \$36.19 for compute and
\$7.67 for network IO. \sys increases the compute cost to \$43.12. \Sys's
network cost depends on the parameter exchange policy, ranging from \$7.67
if parameters are exchanged using the average-iteration-length policy to
\$23.00 for the $1/3$-average policy.  Note, however, network bandwidth
within a single data center is free, and model training would likely
occur within a single data center.  Our experiments with CIFAR show that
more frequent updates consume network bandwidth but provide better models.

\section{Related Work}
\label{s:related}

\paragraphbe{Secure ML environments.}
Ohrimenko et al.\ describe an SGX-based system for multi-party machine
learning on an untrusted platform~\cite{ohrimenko2016}.  They focus
on \emph{collaborative learning}, as opposed to \sys, which focuses on
\emph{outsourced learning} and, in particular, machine learning provided
as a service by a cloud operator.

The critical distinction between the two scenarios is that
in collaborative learning, the model architecture and learning
algorithms are \emph{public}, whereas in ML-as-a-service, they are
\emph{secret} and proprietary to the service provider.  A significant
advantage of~\cite{ohrimenko2016} is that their learning algorithms
are data-oblivious and thus secure against page-fault side channels.
Obliviousness, however, must be verified by the clients, thus the
entire codebase inside the SGX enclave must be public, including any
data-dependent model design choices.  Therefore, \cite{ohrimenko2016}
is unsuitable for ML-as-a-service, where the model is chosen adaptively
based on the client's data and task, but (in most existing ML services)
remains hidden from the client.  By contrast, \sys (1) lets training SGX
enclaves execute untrusted service provider code to adaptively define
the model, but (2) sandboxes this code to prevent it from leaking the
data outside the enclave.

CQSTR~\cite{cqstr2016} lets a \emph{trusted} platform operator confine
untrusted machine learning code so that it can be securely applied
to user data.  By contrast, \sys protects user data from an untrusted
platform operator.

\paragraphbe{Cryptographically protected ML.}
Many papers describe how to use cryptographic techniques,
including secure multi-party computation, to learn and
apply relatively simple classifiers without decrypting the
data~\cite{lindell2002,brickell2007,vaidya2008,brickell2009,bost2015ml}.
It is not clear whether and how these techniques can be applied to modern
ML problems, such as training deep neural networks on ImageNet-scale
datasets, without a prohibitive performance penalty.

More recent work~\cite{dowlin16icml,miniONN} demonstrated how to
efficiently \emph{apply} neural networks to encrypted data.  As far as
we know, today there are no practical techniques for
\emph{training} deep neural networks on encrypted data.

\paragraphbe{Leakage of training data from ML models.}
Overfitted machine learning models can be vulnerable to
``membership inference'': an adversary can infer, with black-box
access to the model, whether a given input was used to train the
model~\cite{shokri2017membership}.  A malicious training algorithm can
construct a model that leaks significant parts of the training dataset
in response to certain queries~\cite{song2017ccs}.

Information leakage from the model's training dataset can be mitigated
by differentially private learning algorithms~\cite{abadi2016ccs,
papernot2017}.  These algorithms produce the same model with approximately
the same probability if a particular input was included in the
training dataset or not.

In this paper, we assume that the owner of the training dataset and
the user of the resulting model are the same, and that the model is
not exposed to other parties (including the operator of the service
that trained the model).  This assumption holds for many common uses
of ML-as-a-service.  In this scenario, there are no adversaries who have
query access to the trained model.

\paragraphbe{SGX-protected execution enviroments.}
Several recently proposed systems aim to protect applications from an
untrusted platform.  Haven~\cite{haven14osdi}, SCONE~\cite{scone16}, and
Graphene-SGX~\cite{tsai17atc} provide an environment to support unmodified
legacy applications.  VC3~\cite{vc3} and Opaque~\cite{zheng17nsdi}
provide SGX-protected data processing platforms.  All of these systems
assume that all of the code inside the enclave is trusted. \Sys protects
user data from untrusted code.

\paragraphbe{Side-channel attacks on SGX.}
As discussed in \S\ref{s:sgxbroke}, side channels can subvert
the protections provided by SGX.  Side channels based on page
faults~\cite{xu15oakland,wang17ccs} enable the platform to
extract private data by observing page-granularity memory accesses.
Physical monitoring of the bus gives an attacker access to word-granularity
memory accesses.  Existing side channels on modern processors (e.g.,
cache~\cite{brasser17woot}, power, and time) apply to enclave-protected
code as well~\cite{sgx-side-channels}.

Processor monitoring units (PMUs) record information about execution and
can leak information about enclave code if they are not cleared.
Lee et al.
demonstrate an attack using the branch history~\cite{branchShadowing}.
The processor's uncore counters (e.g., L3 events) are
enabled during enclave execution~\cite{costan16sgxbook} and could
possibly be used to construct a side channel.

\section{Conclusion}
We presented \sys, a new system for privacy-preserving outsourced
machine learning (ML).  \sys enables data holders to use ML-as-a-service
without disclosing their data to the service providers, yet\textemdash in
keeping with the current practice of these services\textemdash does not
require that the providers make their models, configuration parameters,
and training algorithms public.

\newpage
\bibliographystyle{plain}
\bibliography{bibliography}

\end{document}